\documentclass[sigconf]{acmart}
\AtBeginDocument{%
  }

\setcopyright{acmlicensed}
\copyrightyear{2026}
\acmYear{2026}
\acmConference[SIGIR '26]{Make sure to enter the correct
  conference title from your rights confirmation email}{June 03--05,
  2018}{Melbourne, Australia}
\makeatletter
\gdef\@copyrightpermission{
   \begin{minipage}{0.3\columnwidth}
     \href{https://creativecommons.org/licenses/by-nc-sa/4.0/}{\includegraphics[width=0.90\textwidth]{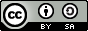}}
   \end{minipage}\hfill
   \begin{minipage}{0.7\columnwidth}
     \href{https://creativecommons.org/licenses/by-nc-sa/4.0/}{This work is licensed under a Creative Commons Attribution-ShareAlike International 4.0 License.}
   \end{minipage}
   \vspace{5pt}
}
\makeatother

\begin{document}

\title{The LLM Effect on IR Benchmarks: A Meta-Analysis of Effectiveness, Baselines, and Contamination}

\author{Moritz Staudinger}
\orcid{0000-0002-5164-2690}
\affiliation{%
  \institution{TU Wien}
  \city{Vienna}
  \country{Austria}
}
\email{moritz.staudinger@tuwien.ac.at}

\author{Wojciech Kusa}
\orcid{0000-0003-4420-4147}
\affiliation{%
  \institution{NASK}
  \city{Warsaw}
  \country{Poland}}
  
\email{wojciech.kusa@tuwien.ac.at}
\author{Allan Hanbury}
\orcid{0000-0002-7149-5843}
\affiliation{%
  \institution{TU Wien}
  \city{Vienna}
  \country{Austria}}
\email{allan.hanbury@tuwien.ac.at}
\renewcommand{\shortauthors}{Staudinger et al.}

\begin{abstract}
Benchmark collections have long enabled controlled comparison and cumulative progress in Information Retrieval (IR). However, prior meta-analyses have shown that reported effectiveness gains often fail to accumulate, in part due to the use of weak or outdated baselines. While large language models are increasingly used in retrieval pipelines, their impact on established IR benchmarks has not been systematically analyzed. In this study, we analyze 143 publications reporting results on the TREC Robust04 collection and the TREC Deep Learning 2020 (DL20) passage retrieval benchmark to examine longitudinal trends in retrieval effectiveness and baseline strength. 
We observe what we term an \emph{LLM effect}: recent systems incorporating LLM components achieve 8.8\% higher nDCG@10 on DL20 compared to the best result from TREC 2020 and approximately 20\% higher on Robust04 since 2023. However, adapting a data contamination detection approach to reranking reveals measurable contamination in both benchmarks. While excluding contaminated topics reduces effectiveness, confidence intervals remain wide, making it difficult to determine whether the LLM effect reflects genuine methodological advances or memorization from pretraining data.
\end{abstract}



\maketitle

\section{Introduction}

Information Retrieval (IR) has a long-standing tradition of using shared benchmark datasets to enable controlled comparison of models and methodologies. Since the inception of the Text Retrieval Conference (TREC) in 1992~\cite{DBLP:conf/trec/1992}, standardized test collections with human relevance judgments have served as the primary instrument for measuring progress in retrieval effectiveness. %

Over time, both the scale of the research community and the methodological landscape of IR have changed substantially. In the past decade alone, publication volume at major venues such as SIGIR has grown more than threefold (from 70 full papers in 2015 to 239 in 2025), coinciding with the widespread adoption of neural retrieval architectures. While these developments suggest rapid innovation, multiple meta-analyses have raised concerns about whether reported improvements reliably translate into genuine advances in retrieval effectiveness. Studies by \citet{armstrong_improvements_2009} and \citet{yang_critically_2019} show that results are frequently compared against weak or outdated baselines, inflating perceived progress, while improvements often fail to accumulate when evaluated against strong reference systems~\cite{kharazmi_examining_2016}. As a result, benchmark-driven evaluation may not always reflect real methodological progress.

The recent emergence of large language models (LLMs) introduces a new inflection point in IR research. LLMs are now routinely used for query expansion~\cite{kim2025_queryexpansion}, knowledge distillation~\cite{schlatt_Setencoder2025}, and reranking~\cite{abdallah2025rankifycomprehensivepythontoolkit} among other tasks. Unlike earlier neural architectures, LLMs provide strong zero-shot and few-shot capabilities, blur the boundary between retrieval and generation, and introduce new sources of inductive bias through large-scale pretraining. Consequently, it remains unclear to what extent reported effectiveness gains on standard IR benchmarks reflect genuine methodological advances or follow earlier patterns of non-cumulative progress.

In this work, we examine whether the adoption of LLM-based components has altered longitudinal trends in retrieval effectiveness on established IR benchmarks. Focusing on the TREC Robust04 collection and the TREC Deep Learning 2020 Passage Retrieval benchmark (DL20), we address three questions: (1) whether reported effectiveness gains accumulate relative to strong baselines; (2) how evaluation metrics and practices have evolved alongside neural and LLM-based methods; and (3) whether potential data contamination can explain observed gains.
To address these questions, we build on prior studies spanning 2005–2018 and conduct a longitudinal meta-analysis of 143 publications on Robust04 (2005-2025) and DL20 (2020–2025), analyze shifts in evaluation practice and their implications for comparability, and adapt the Data Contamination Quiz~(DCQ)~\cite{dcq_2025} to the reranking setting to assess whether reported LLM-based gains can be attributed to training--test overlap.

\section{Meta-Analysis}\label{sec:meta}
To investigate whether LLMs have led to measurable effectiveness gains on established IR benchmarks, we conducted a meta-analysis of reported experimental results on the Robust04 and DL20 datasets. Our objective is to trace longitudinal trends in reported retrieval effectiveness, assess baseline strength, and determine whether systems incorporating LLM components are associated with higher effectiveness.

\begin{figure*}[t]
  \centering
  \includegraphics[width=0.95\textwidth]{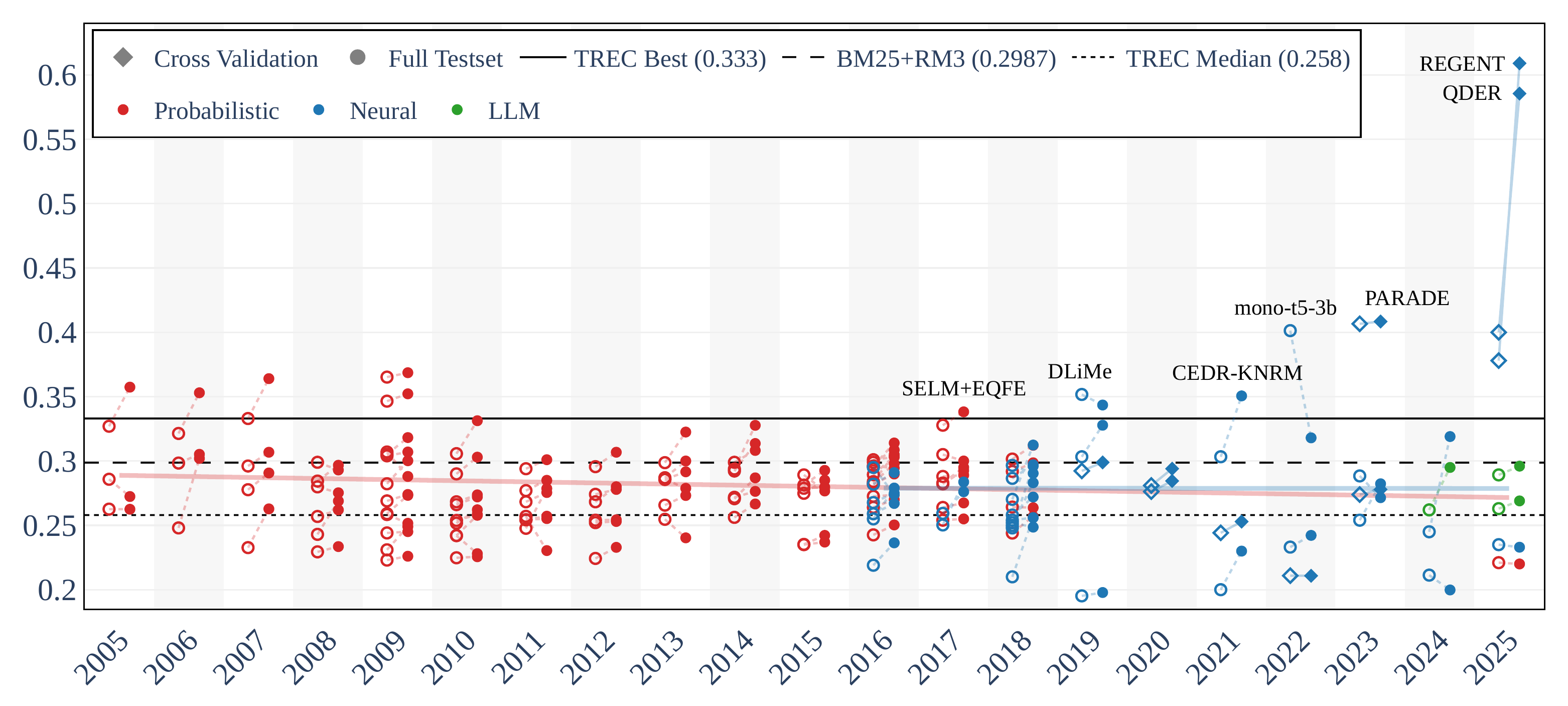}
  \caption{Robust04 MAP results between 2005 and 2025. Regression lines show trends based on best reported results. Empty circles and diamonds show reported baselines, full circles and diamonds best reported results.}
  \label{fig:rob04_map}
\end{figure*}

We performed a systematic literature search on the ACM Digital Library\footnote{\url{https://dl.acm.org/}} to identify publications reporting results on Robust04 and DL20. Separate keyword-based queries were issued for each collection.\footnote{\textbf{Robust04 query}: [Full Text: "robust04"] OR [[Full Text: "robust 04"] AND [Full Text: "trec robust04"]] OR [Full Text: "trec robust"] OR [[Full Text: "trec"] AND [Full Text: "disk 4"]] AND [E-Publication Date: (01/01/2019--12/31/2025)]. \textbf{DL20 query}: [Full Text: "dl20"] OR [Full Text: "deep learning 2020"] OR [Full Text: "dl-20"] OR [Full Text: "deep learning 20"] AND [E-Publication Date: (01/01/2020--12/31/2025)].}
We restricted the analysis to publications from 2019–2025 for Robust04, as earlier work covered up to 2019~\cite{armstrong_improvements_2009, yang_critically_2019}, and to 2020–2025 for DL20, which was introduced in late 2020.

\begin{figure}[!htbp]
  \centering
  \includegraphics[width=\columnwidth]{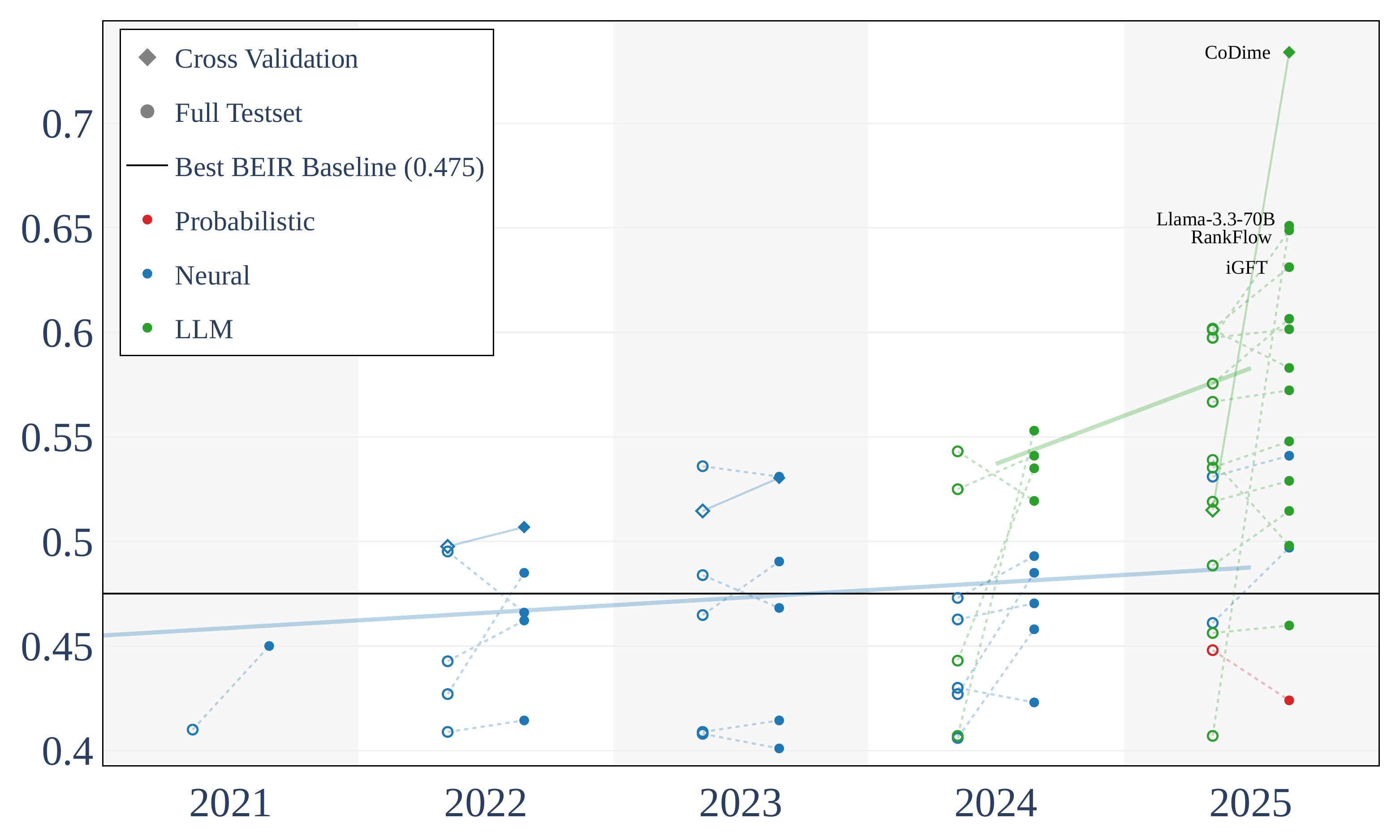}
  \caption{Robust04 nDCG@10 results between 2021 and 2025. Empty circles show baselines, full circles best results.}
  \label{fig:rob04_ndcg}
\end{figure}

\begin{figure}[!htbp]
  \centering
  \includegraphics[width=\columnwidth, trim=0cm 0cm 0cm 0cm, clip]{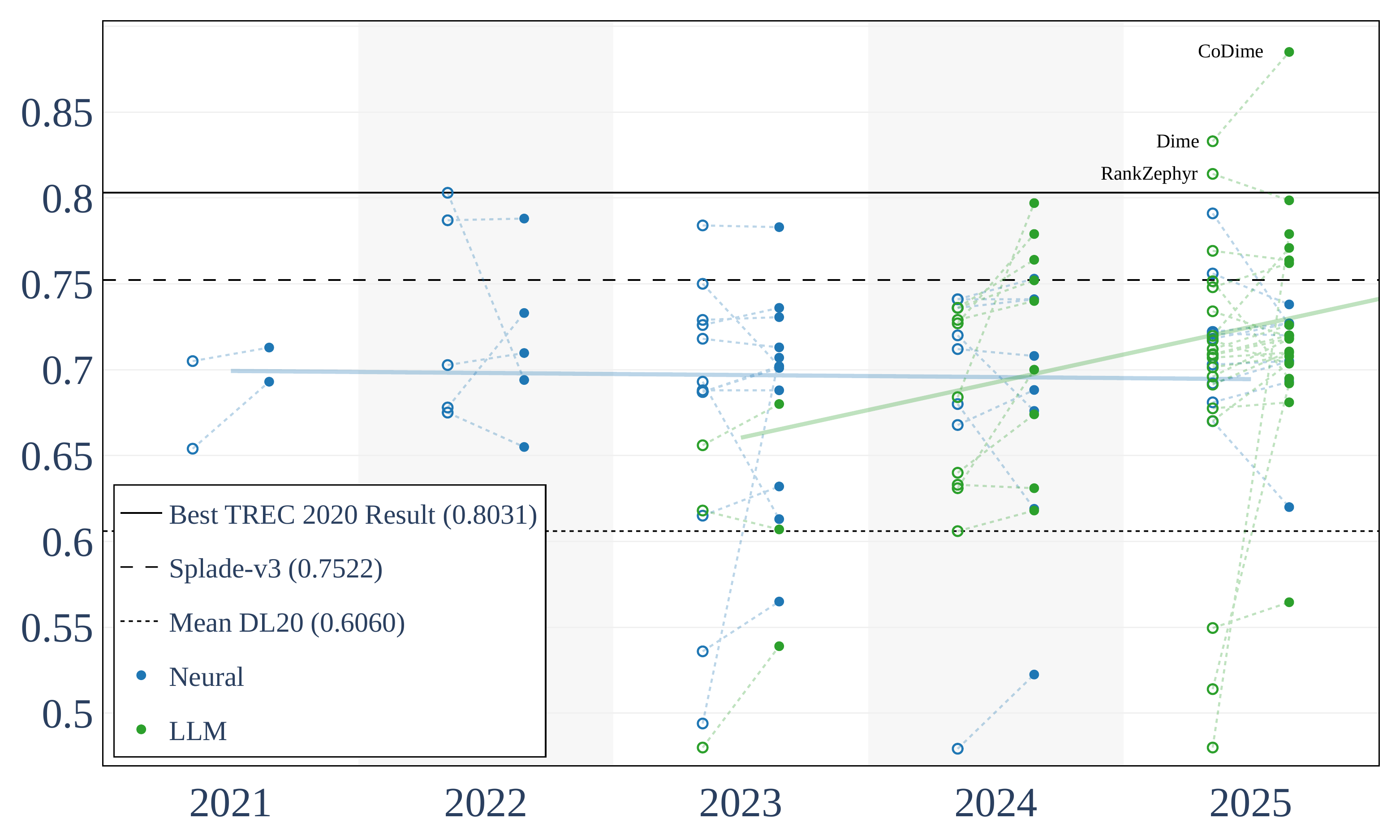}
  \caption{TREC DL2020 nDCG@10 results between 2021 and 2025. Empty circles show baselines, full circles best results.}
  \label{fig:dl20_ndcg}
\end{figure}

The screening process was conducted in multiple stages. First, each retrieved paper was checked for an experimental evaluation on either the Robust04 or the DL20 dataset. Papers not reporting experiments on at least one of these collections were excluded. For each remaining paper, we extracted two results: (1) the highest effectiveness reported for the proposed model and (2) the strongest baseline from prior work, used by the authors. When authors reimplemented existing systems without proposing new models, these results were also included. The screening and extraction process yielded results from $72$ Robust04 papers and $71$ DL20 papers.
We manually compiled the strongest reported baselines and best-performing models across the reported retrieval metrics, including Mean Average Precision (MAP) and Normalized Discounted Cumulative Gain (nDCG). We categorize models into three groups: \emph{probabilistic} (traditional models like BM25), \emph{neural} (deep learning models without LLM components), and \emph{LLM} (models with more than 7B parameters or explicitly containing ``LLM'' in their name). These results form the basis for the longitudinal analysis shown in Figure~\ref{fig:rob04_map}, \ref{fig:rob04_ndcg} and \ref{fig:dl20_ndcg}.

Figure~\ref{fig:rob04_map} extends the analyses of \citet{armstrong_improvements_2009} and \citet{yang_critically_2019}, presenting MAP results from 2005–2025: the earlier period (2005–2018) follows the coverage of these prior studies, while we add results from 2019–2025. Regression lines indicate trends for each model category. We observe the improvements noted by \citet{lin_neural_2021} in 2019, with several recent models now outperforming the original TREC best result from 2004 (0.333, solid line), including monoT5-3B~\cite{nogueira-etal-2020-document}.

Many recent models perform cross-validation (CV, indicated by diamonds)\footnote{Only from 2019 onwards} on Robust04 to enable training on the dataset. While standardized CV splits exist, such as those proposed by \citet{huston_croft_2014}, not all researchers use them. The best-performing models, QDER~\cite{qder_2025} (MAP of 0.5855) and REGENT~\cite{regent_2025} (MAP of 0.609), both employ a custom five-fold CV and average results across folds. Because CV-based results are not directly comparable to full test-set evaluations, we visualize them separately and interpret trends cautiously. When excluding these two outliers, neural model performance appears largely flat over time, although several recent systems surpass the original TREC best model. Only three LLM-based models report MAP on Robust04, limiting trend analysis for this metric.

LLM-based models more frequently report nDCG@10, which has become the most common metric in recent Robust04 evaluation. As shown in Figure~\ref{fig:rob04_ndcg}, we observe steadily increasing performance over the years, with most newly proposed models in 2024 and 2025 using LLMs for parts of their retrieval pipeline. The best-performing system (CoDime~\cite{codime2025}) achieves nDCG@10 of $0.734$, representing an improvement of 20\% to the best result of 2023 ($0.536$) %

A similar trend can be observed in Figure~\ref{fig:dl20_ndcg} for nDCG@10 on the DL20 dataset. While many LLM-based models in 2025 still perform below the best TREC result from 2020 (0.8031, solid line)~\cite{pash2020} and even SpladeV3~\cite{lassance2024spladev3newbaselinessplade} (0.7522, dashed line, best available prebuilt index in Anserini~\cite{lin_toward_2016}), five models also outperform SpladeV3 and two TREC. The best-performing model in this group is CoDIME~\cite{codime2025}, which reaches an nDCG@10 of $0.885$.

Taken together, the results across Robust04 and DL20 indicate a general upward trend in reported neural and LLM retrieval effectiveness over time. While performance gains in systems using LLMs are not uniform across metrics or evaluation setups, and are influenced by shifts in reporting practices and baselines, our analysis reveals consistent improvements: a 20\% higher nDCG@10 on Robust04 since 2023, a 8.8\% higher on DL20 compared to the best TREC2020 result. Recent systems -- many incorporating LLM components -- tend to achieve higher effectiveness than earlier approaches. We refer to this descriptive pattern as the \emph{LLM effect}.%

\section{Data Contamination}
As many publicly available LLMs are pretrained on partially undisclosed corpora, we additionally investigate the potential impact of data contamination on benchmark evaluation. While data contamination is a well-known issue in Natural Language Processing, its extent and implications for IR evaluation remain underexplored. Prior work has examined contamination in ranking distillation~\cite{kalal_training_2024} and zero-shot retrieval settings~\cite{frobe_how_2022}, but systematic analyses on publicly available models are limited. %

To estimate contamination levels on DL20 and Robust04, we adapt the DCQ proposed by \citet{dcq_2025}. In the original setup, an LLM is repeatedly asked to identify an original text fragment from a set of four similar variants to detect verbatim memorization while mitigating positional bias. We adapt this approach for reranking settings: for each topic, we sample five relevant passages from Robust04 (1,149 query--document pairs) and DL20 (266 pairs). For each passage, we generate four paraphrased variants using \textit{gemini-2.5-flash}, creating a four-option multiple-choice question, where one option is replaced with the original passage and the other options are semantically equivalent paraphrases conveying identical information. We then apply the DCQ procedure, modifying the prompt to include the associated topic and asking the model to identify the original text. Since all variants are equally relevant to the topic, correct identification indicates memorization rather than relevance judgement ability.

Table~\ref{tab:reranking_comparison} reports the resulting contamination estimates for two widely used LLM-based rerankers. On DL20, RankZephyr exhibits contamination levels between $26.12\%$ and $31.95\%$, while RankGPT (\textit{gpt-4o-mini}) reaches approximately $41\%$\footnote{Chance-level guessing is accounted for following the DCQ methodology.}. Even Robust04, despite restricted-access, shows 12--21\% contamination,likely reflecting the inclusion of public newswire sources in pretraining corpora.

To assess the effect of contamination on retrieval effectiveness, we conduct a reranking experiment with RankZephyr and RankGPT on both datasets. For Robust04 and DL20, we retrieve candidate passages using Splade++\footnote{BM25+RM3 runs are available in our repository} and rerank the top 100 results per topic. Each model is evaluated once on the full topic set (PC= potentially contaminated) and once after excluding all topics for which the model correctly identified at least one relevant passage in the DCQ (NC= non-contaminated). While this conservative filtering substantially reduces evaluated topics - retaining only $11\%$ of DL20 and $42\%$ of Robust04 topics for RankZephyr - it provides an upper bound estimate of potential impact of contamination.

As shown in Table~\ref{tab:reranking_comparison}, excluding potentially contaminated topics generally reduces effectiveness, but stays within the bootstrapped confidence intervals (CI, 10{,}000 bootstrapped samples from the PC-NC topics). On DL20, filtering contaminated topics reduces nDCG@10 by approximately $14\%$ for RankZephyr and $13\%$ for RankGPT, although this comparison is based on a small number of remaining topics, and therefore the CI is large. MAP increases by $3\%$ for RankZephyr but drops $19\%$ for RankGPT. On Robust04, the effect is smaller, also due to the larger testsets: RankZephyr’s nDCG@10 decreases slightly from 0.5522 to 0.5440 after filtering, while RankGPT drops more noticeably from 0.5306 to 0.4990; MAP shows a noticeable decrease to $0.1774$, lying outside of the CIs.

Overall, contamination appears measurable on both benchmarks, but its impact on reported effectiveness remains difficult to quantify reliably in IR settings. The small number of non-contaminated topics after filtering, together with uncertainty in adapting DCQ to reranking, makes the resulting effectiveness differences inconclusive. While contamination may contribute to observed gains, the current analysis does not allow a clear attribution of improvements to either methodological advances or memorization effects. We therefore interpret these findings cautiously and view current contamination detection methods as insufficient for reliably estimating its impact on IR benchmark evaluation.

\begin{table}[t]
\caption{Comparison of RankZephyr and RankGPT on DL2020 Passage Retrieval and Robust04. 
PC = full (potentially contaminated) test set, NC = non-contaminated topics only. Scores outside of CIs are marked with \textsuperscript{\textdagger}.}
\centering
\small
\setlength{\tabcolsep}{6pt}

\begin{tabular}{lcccc}
\toprule
 & \multicolumn{2}{c}{\textbf{RankZephyr}} & \multicolumn{2}{c}{\textbf{RankGPT}} \\
 & \textbf{PC} & \textbf{NC} & \textbf{PC} & \textbf{NC} \\
\midrule

\multicolumn{5}{l}{\textbf{DL2020 Passage Retrieval}} \\
\hphantom{xx}Input pairs & \multicolumn{4}{c}{266} \\
\hphantom{xx}Contamination level 
& \multicolumn{2}{c}{[26.12\%, 31.95\%]} 
& \multicolumn{2}{c}{[40.98\%, 41.73\%]} \\
\midrule
Topics evaluated & 54 & 6 & 54 & 4 \\
\addlinespace[2pt]
Splade++ nDCG@10 & 0.7849 & 0.6437\textsuperscript{\textdagger} & 0.7838 & 0.6502 \\
\hphantom{xx}\textit{CI$_{95}$ Splade++ nDCG@10} 
& \multicolumn{2}{c}{[0.6741, 0.9163]}
& \multicolumn{2}{c}{[0.5756, 0.9271]} \\
\addlinespace[2pt]
Splade++ MAP     & 0.5731 & 0.6022 & 0.5279 & 0.3522 \\
\hphantom{xx}\textit{CI$_{95}$ Splade++ MAP} 
& \multicolumn{2}{c}{[0.3832, 0.7598]}
& \multicolumn{2}{c}{[0.3115, 0.7533]} \\
\addlinespace[2pt]
\midrule \midrule
\multicolumn{5}{l}{\textbf{Robust04}} \\
\hphantom{xx}Input pairs & \multicolumn{4}{c}{1149} \\
\hphantom{xx}Contamination level 
& \multicolumn{2}{c}{[12.36\%, 16.10\%]} 
& \multicolumn{2}{c}{[20.14\%, 20.97\%]} \\
\midrule
Topics evaluated & 250 & 105 & 250 & 57 \\
\addlinespace[2pt]
Splade++ nDCG@10 & 0.5522 & 0.5440 & 0.5328 & 0.4990 \\
\hphantom{xx}\textit{CI$_{95}$ Splade++ nDCG@10} 
& \multicolumn{2}{c}{[0.5266, 0.5907]}
& \multicolumn{2}{c}{[0.4769, 0.6088]} \\
\addlinespace[2pt]
Splade++ MAP     & 0.2589 & 0.2464 & 0.2284 & 0.1774\textsuperscript{\textdagger} \\
\hphantom{xx}\textit{CI$_{95}$ Splade++ MAP} 
& \multicolumn{2}{c}{[0.2456, 0.2892]}
& \multicolumn{2}{c}{[0.2027, 0.2865]} \\
\addlinespace[2pt]

\bottomrule
\end{tabular}
\label{tab:reranking_comparison}
\end{table}

\section{Discussion}
\paragraph{Effectiveness Trends}
We observe upward trends in reported neural and LLM effectiveness on both benchmarks, supporting our characterization of an \emph{LLM effect}. These gains are modest but consistent across evaluation settings. However, not all contributions primarily target raw effectiveness—many focus on efficiency, robustness, or explainability—partially explaining why improvements accumulate gradually rather than rapidly.

\paragraph{Evaluation Practice Shifts}
A critical finding concerns metric diversity, particularly on Robust04, where 72 publications report 19 different metrics. The most frequent is nDCG@10 (40 publications), followed by MAP (27) and nDCG@20 (26). In contrast, DL20 evaluation is more standardized: 65 of 71 publications report nDCG@10. This heterogeneity complicates longitudinal comparison, as improvements may not be directly comparable across studies using different metrics or cutoff values.

The shift from MAP to nDCG@10 fundamentally changes what constitutes "progress." MAP emphasizes overall ranking quality while nDCG@10 emphasizes top-rank precision. The flatter MAP trend (Figure~\ref{fig:rob04_map}) versus upward nDCG@10 trend (Figure~\ref{fig:rob04_ndcg}) likely reflects nDCG's higher sensitivity to early-rank improvements where recent reranking-focused systems excel. The dominance of nDCG@10 coincides with BEIR's 2021 release~\cite{thakur_beir_2021}, which standardized evaluation across 18 datasets including Robust04. This temporal correlation between metric shifts and LLM adoption raises an important question: does the observed LLM effect partially reflect metric selection rather than purely methodological advances?

Our literature review further reveals that the TREC Deep Learning Document Retrieval collection is rarely used: only six publications report document-level effectiveness, with the majority evaluating passage retrieval.

\paragraph{Limitations}
Restricting our search to ACM Digital Library excludes NLP and ML venues (EMNLP, ACL or NeurIPS) and Springer proceedings (ECIR, IRJ), potentially underestimating baselines. We maintain this restriction for consistency with prior meta-analyses~\cite{armstrong_improvements_2009, yang_critically_2019}. Second, keyword-based search may miss relevant publications due to imperfect query formulation or indexing delays. Third, dataset ambiguity complicates comparison: different MS MARCO versions yield substantially different scores~\cite{lassance_tale_2023}, and heterogeneous cross-validation setups on Robust04 limit comparability.

Finally, while our contamination analysis reveals measurable leakage (26--41\% on DL20, 12--21\% on Robust04) with detectable effectiveness impacts, conservative filtering yields small samples and wide confidence intervals that prevent definitive quantification. This introduces uncertainty when attributing recent effectiveness gains to methodological advances versus memorization effects.

These findings suggest the IR community should establish consensus on standard metrics or require reporting multiple complementary metrics. Future benchmark evaluations should include contamination detection as standard practice when claiming state-of-the-art results with models pretrained on undisclosed corpora.

\section{Conclusion}\label{sec:conclusion}
We examined the \emph{LLM effect} on retrieval evaluation, asking whether LLM adoption has altered effectiveness reporting patterns on established IR benchmarks. Through a longitudinal meta-analysis of 143 publications on Robust04 and TREC Deep Learning 2020 Passage Retrieval, we observed upward trends in reported effectiveness—with recent systems achieving up to 11\% higher nDCG@10—alongside significant shifts in evaluation practice, particularly the transition from MAP to nDCG@10.

Our analysis confirms a descriptively positive LLM effect: recent systems incorporating LLM components achieve higher reported effectiveness than earlier approaches when compared against strong baselines. However, by adapting a data contamination detection approach to reranking, we reveal substantial contamination in both benchmarks (26--41\% on DL20, 12--21\% on Robust04). While contaminated topics show reduced effectiveness when excluded, wide confidence intervals prevent definitive attribution of observed gains to either methodological improvements or memorization effects.

These findings indicate that benchmark-driven IR evaluation captures incremental progress in the LLM era, but within an increasingly heterogeneous evaluation landscape that complicates longitudinal comparison. The temporal correlation between metric shifts and LLM adoption raises an important question: does the observed LLM effect partially reflect changing evaluation priorities rather than purely technical advances? Future work should focus on improving the comparability and verifiability of reported results through automated leaderboards, standardized evaluation protocols that reduce metric heterogeneity, and more robust contamination detection methods. Our code and data are available
\href{https://github.com/MoritzStaudinger/LLM-effect}{\color{blue}{\underline{online}}}.

\bibliographystyle{ACM-Reference-Format}
\bibliography{references, software}

\end{document}